# Optical investigation of magneto-structural phase transition in FeRh

V. Saidl[1], M. Brajer[1], L. Horák[1], H. Reichlová[2], K. Výborný[2], M. Veis[1], T. Janda[1], F. Trojánek[1], I. Fina[3], X. Marti[2], T. Jungwirth[2,4], and P. Němec[1,a)]

[1]*Faculty of Mathematics and Physics, Charles University in Prague, 121 16 Prague, Czech Republic*
[2]*Institute of Physics ASCR, v.v.i., 162 53 Prague, Czech Republic*
[3]*ICN2-Institut Català de Nanociència i Nanotecnologia, Campus Universitat Autònoma de Barcelona, 08193 Bellaterra, Spain*
[4]*School of Physics and Astronomy, University of Nottingham, NG7 2RD Nottingham, United Kingdom*

Magneto-structural phase transition in FeRh epitaxial layers was studied optically. It is shown that the transition between the low-temperature antiferromagnetic phase and the high-temperature ferromagnetic phase is accompanied by a rather large change of the optical response in the visible and near infrared spectral ranges. This phenomenon was used to measure the phase transition temperature in FeRh films with thicknesses from 6 to 100 nm and it was observed that the hysteretic transition region broadens significantly in the thinner films.

The near-equiatomic FeRh alloy undergoes a first-order magneto-structural transition from an antiferromagnetic (AF) to ferromagnetic (FM) phase around 380 K. Even though this discovery is more than seventy years old[1] it still attracts a significant attention nowadays. One of the reasons is that this transition occurs close to the room temperature which makes this material rather appealing for applications. For example, in 2003 Tiele *et al.* used exchange-coupled FePt/FeRh thin films to lower the coercive field and improve the long-time stability of medium used in thermally-assisted magnetic recording.[2] In 2014 Marti *et al.* reported a room-temperature FeRh antiferromagnetic memory resistor,[3] which demonstrated the feasibility of the antiferromagnetic spintronics concept.[4] In the same year Cherifi *et al.* reported an electric-field control of the transition temperature in FeRh film grown on the ferroelectric substrate $BaTiO_3$.[5] In 2015 Lee *et al.* used the strain-mediated change in the relative proportions of AF and FM phases to construct the memory device based on FeRh/PMN-PT.[6] Moreover, it was shown that the change of the magnetization order can be rather fast - AF to FM transition can be induced by laser pulses on the picosecond time scale.[7]

The research of FeRh was originally conducted on bulk samples[1,8-11] but more recently, due to the envisioned applications,[2,3,6] it concentrates on the research of thin films. The desired AF phase is present only for a rather narrow interval of the Fe-to-Rh concentrations,[9] which are achieved experimentally by a variation of the deposition conditions[12-14] and by a post-preparation heat treatment.[15] Consequently, the growth of high quality thin films of FeRh is a rather complex task which requires a lot of optimization of the preparation procedure and a deposition of large series of samples.[12-15] The magnetic phase transition in the prepared films is usually studied by magnetometry, for example by superconducting quantum interference device (SQUID), where not only the phase transition temperature but also the remaining uncompensated magnetic moment in the low-temperature AF phase is measured.[13,14] The disadvantage of magnetometry is that the magnetic properties

---

[a)] Electronic mail: nemec@karlov.mff.cuni.cz

are averaged over the whole sample and that it is necessary to subtract the contribution of the substrate from the measured data. Moreover, the utilization of SQUID for a characterization of large series of samples[12-15] is also not ideal due to the rather high running costs connected with a consumption of liquid helium even for measurements above the room temperature. In fact, in a typical SQUID machine the temperature can be usually increased only up to 400 K that is quite often not sufficient to fully convert AF to FM phase and, therefore, both the transition temperature and the magnetization in FM phase are not measured precisely in this experiment.[13,14] Alternatively, the transition temperatures can be determined from the resistivity measurements[11] but these experiments are again not ideal for a routine testing of a large set of samples. Better suited for this purpose are measurements of a magneto-optical response of the FeRh samples where a rotation of light polarization plane due to polar Kerr effect was shown to reveal clearly the transition temperature.[15]

In this paper we show that AF to FM phase transition can be also studied by the most straightforward optical experiment where the sample transmission and/or reflectivity is measured during the temperature change. We observe that the measured change of the optical properties of FeRh is most significantly pronounced in the near infrared spectral range. Therefore, the red/infrared laser diodes are the best choice for a construction of a very cost-effective characterization experimental setup where the prepared samples can be routinely tested. We also show that these optical measurements can provide information not only about the phase transition temperature but also about the magnetic quality of FeRh samples.

The published results about the experimentally measured optical properties of FeRh are very limited.[8-10] In 1974 Sasovskaja and Noskov reported ellipsometric measurements of optical constants in bulk FeRh in the infrared spectral region and they observed a clear hysteretic behavior of the sample reflectivity when the temperature was changed from 293 to 423 K and from 423 to 293 K, respectively.[8] In contrast, no appreciable change of optical constants at AF-FM phase transition was reported in 1988 by Chen and Lynch (Ref. 9) and in 1995 by Rhee and Lynch (Ref. 10). In fact, it was pointed out in Ref. 9 that these measurements showed that the band structure of FeRh is less affected by AF to FM phase transition than band-structure calculations indicate.

Our FeRh films with thicknesses from 6 nm to 100 nm were grown on double-side polished MgO (001) substrates by dc sputtering. Firstly, substrates were heated up to 550 K in a base pressure of $10^{-8}$ torr. Afterwards, Ar gas was introduced (3 mtorr) and films were grown using 50 W power at a rate 1 nm per 1 min using a stoichiometric FeRh target. Finally, the films were protected by a 1.5 nm Ta cap. After deposition, films were heated up to 500°C at 10°C/min in vacuum and annealed for 1 h and subsequently cooled to room temperature at 10°C/min.

Standard reflectivity ($R = I_R / I_0$) and transmittance ($T = I_T / I_0$) spectra were computed from the spectral dependences of the reflected ($I_R$), transmitted ($I_T$) and incident ($I_0$) light intensity measured by grating spectrographs equipped with CCD (USB2000+, Ocean Optics). The reflectivity spectra were further corrected for the chromatic aberrations in the experimental setup by comparing the obtained reflectivity spectrum for a reference GaAs sample with the theoretical spectrum computed from the tabulated optical constants of GaAs.[16] The transmittance spectra of FeRh films were corrected for the transmittance and the reflectivity of the MgO substrate. No external magnetic field was applied during the optical measurements. The samples were placed in a vacuum on a cold finger of a closed-cycle cryostat (Advanced Research Systems) where the temperature can be changed from 8 to 800 K. Magnetic properties of the samples were measured using SQUID (Quantum Design) with 4 cm long reciprocating sample option. Data measured above 400 K were obtained using a high-temperature insert. The X-ray diffraction (XRD) experiment was carried out with the standard laboratory coplanar diffractometer (X'Pert MRD panalytical) equipped with the high-temperature chamber (Anton Paar DHS1100) at no external magnetic field applied. We measured in the high-resolution mode using the hybrid monochromator in the incident path and the analyser crystal in front of the point detector. The wavelength of the primary beam was Cu-K$_{\alpha 1}$. We measured the *θ-2θ* scans across the symmetric diffraction FeRh (002) for *2θ* in the range 60° – 63°.

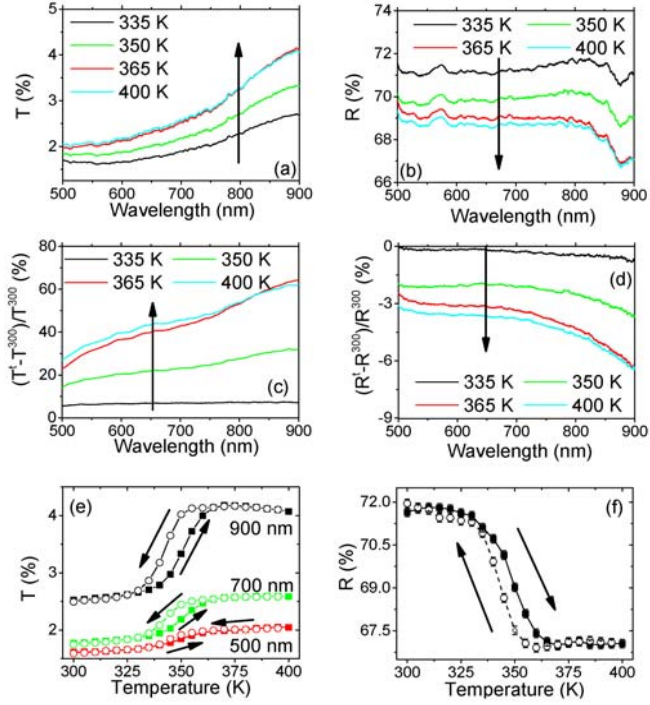

Fig. 1. Temperature dependence of the optical response in 36 nm thick FeRh film. (a) Transmittance (*T*) and (b) reflectivity (*R*) spectra measured for increasing sample temperature. (c) Spectrum of relative transmittance change ($T^t - T^{300}$) / $T^{300}$ computed from the transmittance spectrum measured at given temperature ($T^t$) and at 300 K ($T^{300}$). (d) Spectrum of relative reflectivity change. (e) Temperature dependence of transmittance at 500 nm, 700 nm and 900 nm. (f) Temperature dependence of reflectivity at 900 nm. Arrows indicate the temperature change direction.

In Fig. 1 we show the transmittance and reflectivity spectra that were measured simultaneously during the heating and the successive cooling of a 36 nm thick FeRh film. We observed that the temperature-induced AF to FM phase transition leads to an increase of the sample transmittance [Fig. 1(a)] and to a decrease of the sample reflectivity [Fig. 1(b)]. For example, at 800 nm, the measured transmittance increase of 1% and the reflectivity decrease of 3% indicate the film absorption increase of 2%. From the measurement point-of-view, the magnitude of the optical properties change *relative* to the value measured in the AF phase is important because it illustrates how "difficult" it is to identify the phase transition-induced optical signal change in the as-measured data obtained from detectors (which are always detected with a certain amount of a noise). In Fig. 1(c) and Fig. 1(d) we show the relative change of the film transmittance and reflectivity, respectively. These data illustrate that the magnitude of the optical properties change is the largest in the red and infrared spectral regions but the transmittance change is considerably larger (i.e., easier to detect) than that of the reflectivity. For example, for this particular sample at 800 nm, the relative transmittance and reflectivity changes are about 50% and 5%, respectively. The characteristic temperature of the magneto-structural phase transition ($T_{tr}$) and the width of the hysteretic transition region ($\Delta T_{tr}$) can be determined from the central position and the width of the measured hysteretic optical properties, respectively. For the 36 nm thick film we evaluated $T_{tr}$ = 348 ± 3 K and $\Delta T_{tr}$ = 8 ± 2 K. We note that, in principle, these numbers can be obtained from the transmitted and reflected light of any wavelength. However, in reality only in the transmission geometry any wavelength from the studied spectral range of 500 to 900 nm can be used for this purpose [see Fig. 1(e)]. On the other hand, in the reflection geometry only light in the red and infrared spectral regions provide a reasonably good signal-to-noise ratio in the measured data [see Fig. 1(f)]. These observations have rather strong implications for a practical realization of optical experimental setups where large quantities of prepared samples can be routinely tested during the FeRh films optimization. If relatively thin (up to ≈ 50 nm) FeRh films are prepared on double-side polished transparent substrates basically any available light source can be used to detect the magnetic phase transition in the transmission geometry. For thicker films, the experiment has to be performed in the reflection geometry and the GaAs-based lasers working in the near infrared spectral region, which are fully compatible with silicon detectors, are probably the best choice from the point-of-view of the stability and acquisition price of the optical characterization setup.

To understand the origin of the observed change in transmittance $T$ and reflectivity $R$ upon crossing the FM/AF phase transition, we performed a model calculation of the spectra of $T$ and $R$ using the 4x4 transfer matrix approach for anisotropic multilayers.[17] Our model comprises a multilayer structure (1.5 nm thick Ta capping layer, 36 nm of FeRh and an infinite MgO substrate). Optical parameters (real and imaginary part of permittivity) of Ta and MgO were taken from the literature[18,19] and optical parameters of FeRh were calculated using the full-potential linearized-augmented-plane-wave method (WIEN2k package).[20] All calculations were performed assuming zero temperature and the high (low) temperature phase investigated in our experiments was modelled by assuming the FM (AF) magnetic structure and appropriate lattice constants as described in the Supplementary material.[21] In there, we also discuss the individual effect of the intra- and inter-band contributions to permittivity spectra.

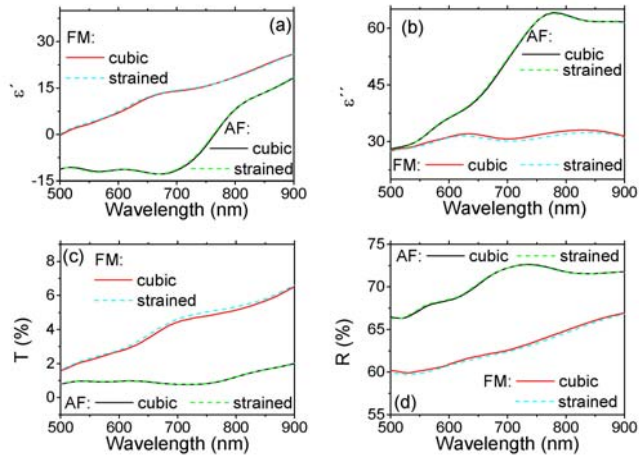

Fig. 2. Calculated spectral dependence of optical parameters of the antiferromagnetic (low-temperature) and ferromagnetic (high-temperature) phase of FeRh for bulk cubic crystal structure and for tetragonal structure in strained thin film. (a) Real ($\varepsilon'$) and (b) imaginary ($\varepsilon''$) part of permittivity of FeRh. Multilayer Ta/FeRh/MgO spectra of transmittance (c) and reflectivity (d).

Real ($\varepsilon'$) and imaginary ($\varepsilon''$) parts of the permittivity are shown in Figs. 2(a) and 2(b), respectively. The obtained transmittance and reflectivity spectra, which are shown in Figs. 2(c) and 2(d), agree qualitatively with the measurements. In particular, they show an increase (decrease) of transmittance (reflectivity) by a few percent upon the transition from the AF to FM phase. In order to explore further the origin of the measured change in optical properties, we compared theoretically the case where both magnetic phases were cubic (bulk-like) to the case of thin film where they were tetragonal due to the MgO substrate-induced strains (MgO has a smaller lattice constant than bulk FeRh).[22] The relevant lattice constants are given in the Supplementary material[21] – here we just note that the substrate-induced strains are smaller in the AF phase than in the FM phase, which has larger lattice constant.[22] We find that the substrate-induced strains do change the permittivity (and consequently the layer transmittance and reflectivity) but the changes are minute. We conclude that the experimentally observed changes of optical properties due to the AF-to-FM transition are not due to the thin film nature of the studied samples and should be equally present also in bulk FeRh samples.

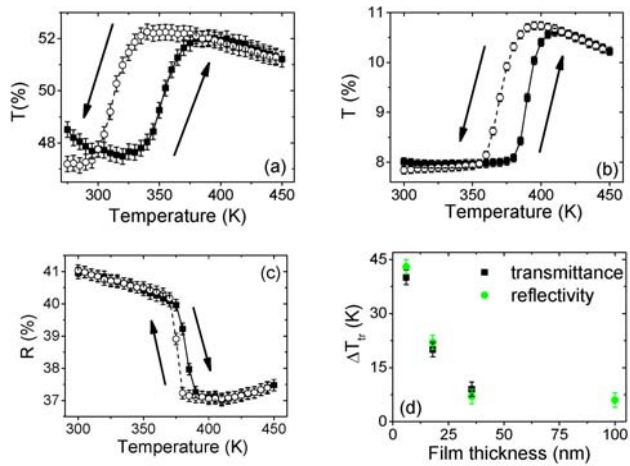

Fig. 3. Change of optical properties of FeRh films with different thicknesses measured at 900 nm due to the temperature-induced phase transition. (a) Transmittance change for 6 nm film, (b) transmittance change for 18 nm film, and (c) reflectivity change for 100 nm film. (d) Dependence of the transition temperature width ($\Delta T_{tr}$) on the film thickness deduced from the transmittance and reflectivity measurements.

In Fig. 3 we show examples of the measured phase transition-induced changes of optical properties of FeRh epitaxial films with different thicknesses. We observe no systematic trend in the thickness dependence of the phase transition temperature ($T_{tr}$): for 6 nm film $T_{tr} = 335$ K, for 18 nm film $T_{tr} = 380$ K, for 36 nm film $T_{tr} = 348$ K [see Fig. 1(e)], and for 100 nm film $T_{tr} = 380$ K. On the other hand, there is a clear monotoneous decrease of the width of the hysteretic transition region ($\Delta T_{tr}$) with the film thickness. (See Fig. 3(d) where the data obtained from the transmittance and reflectivity measurements are shown simultaneously for a direct comparison). Similar trend was reported previously by other groups.[12,13,23] They ascribed it to a reduction of the size of grains forming the film[12] and to a different stability of the AF and FM phase in thin films,[24] which could lead to a lowering of the characteristic temperature of the FM to AF transition during the sample cooling in thinner films while that of the AF to FM transition during the sample heating is not affected significantly by the film thickness.[23] In our case we do not see any signatures of grains in the XRD experiment and also any systematic trend in the thickness dependence of the FM to AF and AF to FM transition temperatures. Therefore, the origin of the observed increase of the width of the hysteretic transition region with the film thickness reduction is not clear at this moment.

The high magnetic and structural quality of all the investigated samples was confirmed by SQUID and XRD measurements. As an example, the temperature dependence of the magnetization in the 36 nm thick film is shown in Fig. 4(a) which is illustrating a negligible moment in the AF phase and the expected[22] magnitude of the moment in the FM phase. In XRD experiments we observed one or two broad diffraction peaks depending on the sample temperature. We associate the peak at a lower diffraction angle with the FM phase and the peak at a higher diffraction angle with the AF phase of FeRh, since these phases differ significantly by the lattice parameter.[22,25] The shape of *both* peaks were modulated by the thickness fringes with the frequency corresponding to the thickness obtained from the X-ray reflectivity measurement. Therefore, we assume that the FeRh layer consists of many neighboring structural domains which are distributed laterally and their vertical dimension is equal to the thickness of the epilayer. Such a model allowed us to fit the measured data with the weighted sum of the squared sinc functions, i.e., as an incoherent sum of the intensity diffracted by two different phases with different lattice parameters. The weighting factors obtained from the fit correspond to the relative volumes of the two phases at a certain temperature. From the diffraction patterns measured at several temperatures during the heating and cooling, we reconstructed the hysteresis loops of the AF and FM volumes, which are shown in Fig. 4(b) for the 36 nm thick film. Consistently with the SQUID measurement, we see that in this sample only a very small fraction of FeRh remains in the FM phase at room temperature.

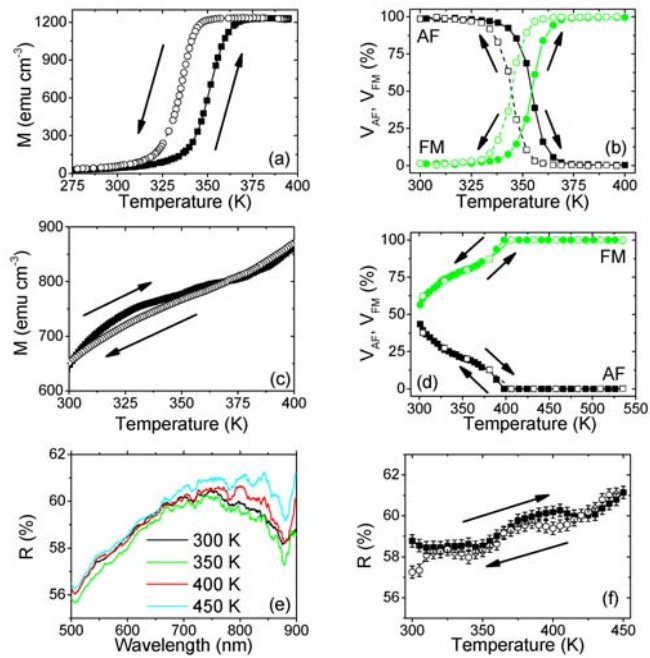

Fig. 4. Comparison of high and low quality FeRh films. Temperature dependence of (a) magnetic moment at an applied field of 1 T and (b) volume of AF and FM phases for high quality 36 nm thick film. Temperature dependence of (c) magnetic moment at an applied field of 0.5 mT and (d) volume of AF and FM phases for low quality 100 nm film. (e) Reflectivity spectra of 100 nm low quality film measured for increasing sample temperature. (f) Temperature dependence of reflectivity at 900 nm of low quality 100 film. Arrows indicate the temperature change direction.

For contrast, we show in Figs. 4(c) and 4(d) SQUID and XRD data obtained in a low quality 100 nm film where a large magnetic moment and nearly the same volume of AF and FM phase were observed at room temperature. This sample was further used to address a question if the low magnetic quality of FeRh film could be identified also in much simpler optical experiment. In Figs. 4(e) and 4(f) we show temperature dependence of the reflectivity that clearly reveals the absence of any hysteretic behavior in the optical response of this sample, which is in a strong contrast to the behavior observed in the high quality film with the same thickness [cf. Fig. 3(c)]. We therefore conclude that the absence of such hysteretic behavior in optical experiments can be regarded as a strong indication of a low magnetic quality of the FeRh film.

In summary, we have shown experimentally that the first-order magneto-structural transition in FeRh is reflected also in the optical response of this material. Our calculations indicate that this modification of optical constants have similar magnitude in thin films and in bulk material. We suggest that this phenomenon can be utilized in a straightforward and cost-effective optical characterization experimental setup where the phase transition temperature and/or quality of prepared FeRh samples can be routinely tested in the near infrared spectral ranges. Moreover, this method of the sample characterization can, in principle, provide even spatially-resolved data with a resolution down to ≈ 1 μm, limited by the size of the focused laser spot.

The authors are indebted to R. Ramesh and S. Salahuddin for preparation of samples. We gratefully acknowledge the assistance of Jakub Zelezny and Martin Ondracek with ab initio calculations. This work was supported by the Grant Agency of the Czech Republic under Grant No. 14-37427G, by EU grant ERC Advanced Grant 268066 - 0MSPIN, and by the Grant Agency of Charles University in Prague Grants no. 1910214 and SVV–2015–260216.

# Optical investigation of magneto-structural phase transition in FeRh

V. Saidl, X. Martí, M. Brajer, L. Horák, H. Reichlová, K. Výborný, M. Veis, T. Janda, F. Trojánek, I. Fina, T. Jungwirth, and P. Němec

Supplemental Material

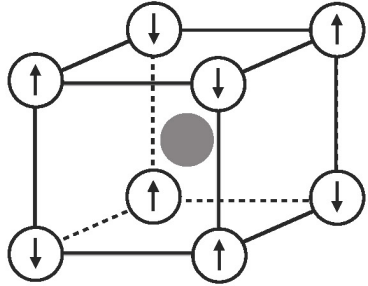

Figure S1: Structure of bulk FeRh. In the AF phase, the Rh atom (grey) has no magnetic moment while the Fe magnetic moments are ordered as indicated. Except for a different lattice constant (see text), the crystal structure remains unchanged in the FM phase; magnetic moment of the Rh atoms is then non-zero and parallel with that of Fe atoms.

## Details of the calculation

Bulk iron rhodium alloy has the cubic structure of caesium chloride with lattice constant $a_{lc}^{\text{FeRh}} = 0.2986$ nm at room temperature [1]. Type-II ordering in the AF phase is shown in Fig. S1. Upon heating, the alloy becomes ferromagnetically ordered and while the lattice constant expands as measured by Ibarra and Algarabel [2], the crystal structure remains unchanged. In accord with these measurements, our calculations of optical properties for the cubic AF (FM) case shown in Figs. 2a,b of the main text assume the lattice constant 0.2986 nm (0.2998 nm).

The substrate (MgO) lattice constant is very well matched to FeRh. At room temperature, $a_{lc}^{\text{MgO}} = 0.4211$ nm [3] implies a mismatch of only $(a_{lc}^{\text{MgO}}/\sqrt{2} - a_{lc}^{\text{FeRh}})/a_{lc}^{\text{FeRh}} \approx -3 \times 10^{-3}$. Assuming that the in-plane lattice constants of FeRh ($a_{xy}^{\text{FeRh}}$) are dictated by the substrate and that the FeRh unit cell volume $(a_{xy}^{\text{FeRh}})^2 a_z^{\text{FeRh}}$ remains unchanged upon this deformation, we arrive at $a_{xy}^{\text{FeRh}} = 0.2978$ nm and $a_z^{\text{FeRh}} = 0.3002$ nm for the strained AF case and $a_{xy}^{\text{FeRh}} = 0.2982$ nm and $a_z^{\text{FeRh}} = 0.3030$ nm for the strained FM case.

The electronic structure of such bulk cubic or tetragonal FeRh crystal was calculated within the local density approximation with zero $U$ and with spin-orbit interaction taken into account. Interband part of AC conductivity $\sigma_{\text{inter}}(\omega)$ was then determined assuming Lorentzian broadening of 0.1 eV.

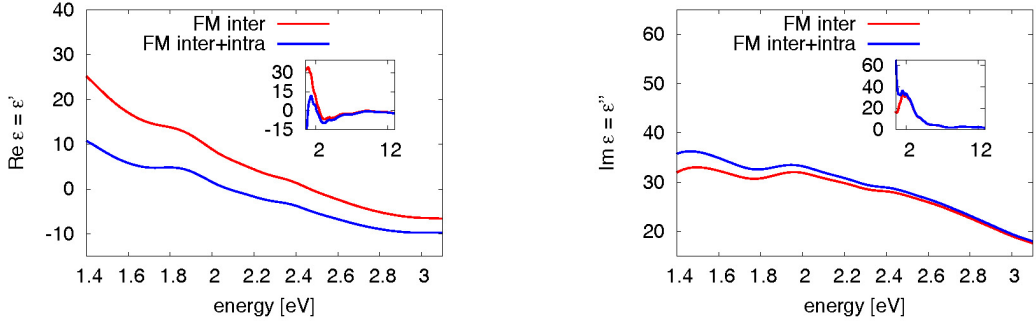

Figure S2: Permittivity $\epsilon(\omega) = \epsilon' + i\epsilon''$ of bulk FeRh in the FM phase. Interband (last term in Eq. (1) only) and total (intraband plus interband) permittivity are shown. Insets show $\epsilon(\omega)$ in a broader spectral range.

## Interband and intraband terms

Permittivity shown in Figs. 2a,b in the main text was calculated using WIEN2k taking into account only interband transitions through $\sigma_{\text{inter}}(\omega)$. The complete expression for $\epsilon(\omega)$ reads

$$\epsilon(\omega)/\epsilon_0 = \epsilon_b - \frac{\omega_p^2}{\omega^2 + 1/\tau^2} + \frac{i\omega_p^2/\omega\tau}{\omega^2 + 1/\tau^2} + \frac{i\sigma_{\text{inter}}(\omega)}{\epsilon_0 \omega} \qquad (1)$$

where $\omega_p$ is the plasma frequency and $\epsilon_b$ is the background relative permittivity stemming from high-energy transitions not accounted for in $\sigma_{\text{inter}}(\omega)$. The plasma frequency given by Eq. (21) in Ref. [4] was determined to be $\hbar\omega_p = 1.8$ eV and 5.5 eV for the AF and FM phase, respectively. The full $\epsilon(\omega)$ as in Eq. (1) calculated by WIEN2k has $\epsilon_b = 1$.

Relaxation time can be estimated from the experimentally known DC conductivity on assumption that $\tau$ is the same for all bands contributing to the Drude formula $\sigma_0 = \omega_p^2 \epsilon \tau$. Taking $\sigma_0 = 11000$ $(\Omega\text{cm})^{-1}$ for both FM and AF phases, total permittivity can be calculated using Eq. (1). Focusing on the experimentally investigated spectral range, $\epsilon(\omega) = \epsilon' + i\epsilon''$ turns out to change only little in the AF phase when intraband transitions are included. In the FM phase, on the other hand, the differences are larger owing to the shorter relaxation times. Spectra of $\epsilon', \epsilon''$ shown in Fig. S2 indicate appreciable intraband corrections mainly to $\epsilon'$ at the largest wavelengths explored.

In spite of their significance on the level of $\epsilon(\omega)$, the inclusion of intraband terms does not change the reflection and transmission spectra qualitatively. The effect of adding intraband terms to $T$ of the FM phase is to add an offset that grows steadily towards the long wavelengths as it is shown in Fig. S3a. This effect is, nevertheless, smaller than the difference to the AF phase. Reflection spectra shown in Fig. S3b display a less clear trend yet it remains true that $R$ decreases by a few percent upon the transition from AF to FM across the whole explored spectral range.

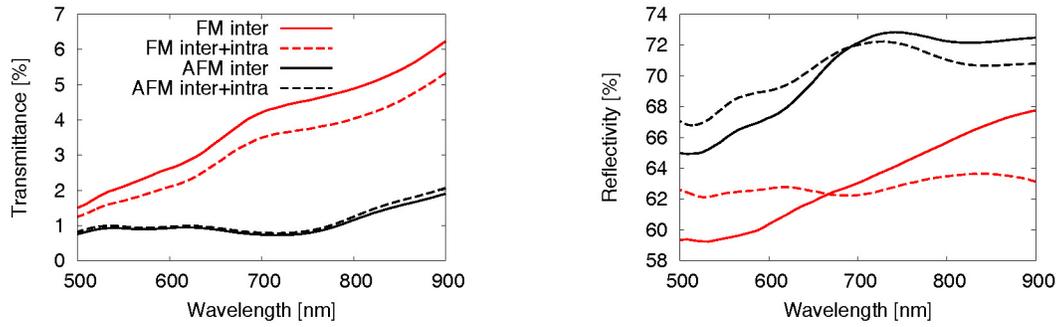

Figure S3: Transmittance and reflectivity for cubic FeRh calculated using permittivity shown in Fig. S2. Inclusion of the intraband contribution (dashed lines) leads only to a quantitative correction of the results shown in Figs. 2c,d of the main text (solid lines).